\documentclass[12pt]{JHEP}

\title{Quantum cosmology, inflationary brane-world creation 
and dS/CFT correspondence}

\author{
Shin'ichi Nojiri \\
Department of Applied Physics \\
National Defence Academy, 
Hashirimizu Yokosuka 239, JAPAN \\
email: nojiri@cc.nda.ac.jp
}
\author{
Sergei D. Odintsov\footnote{On leave 
from Tomsk Pedagogical University, 634041 Tomsk, RUSSIA} \\
Instituto de Fisica de la Universidad de 
Guanajuato \\
Apdo.Postal E-143, 37150 Leon, Gto., MEXICO\\
email: odintsov@ifug5.ugto.mx, odintsov@mail.tomsknet.ru
}
\abstract{
The creation of 4d de Sitter (inflationary) boundary gluing two d5 
de Sitter bulks on the classical as well as on quantum level 
(with account of brane QFT via corresponding trace anomaly 
induced effective action) is discussed. Quantum effects decrease 
the classical de Sitter brane radius or create new de Sitter 
brane with even smaller radius. It is important that brane
CFT may be chosen to be dual to one of 5d de Sitter bulks, 
making the explicit relation of de Sitter brane-world with 
dS/CFT correspondence. In this way, the localization of gravity 
on the brane is shown. Moving 
(time-dependent) de Sitter 
brane in d5 SdS BH is considered. In the special coordinate system 
where brane equations look like quantum-corrected FRW equations 
the comparison with similar brane equations in SAdS BH bulk is done. 
}

\begin{document}
\tolerance=5000
\def\pp{{\, \mid \hskip -1.5mm =}}
\def\cL{{\cal L}}
\def\be{\begin{equation}}
\def\ee{\end{equation}}
\def\bea{\begin{eqnarray}}
\def\eea{\end{eqnarray}}
\def\beaa{\begin{eqnarray*}}
\def\eeaa{\end{eqnarray*}}
\def\tr{{\rm tr}\, }
\def\nn{\nonumber \\}
\def\e{{\rm e}}
\def\D{{D \hskip -3mm /\,}}

\def\SEH{S_{\rm EH}}
\def\SGH{S_{\rm GH}}
\def\AdS5{{{\rm AdS}_5}}
\def\S4{{{\rm S}_4}}
\def\gfv{{g_{(5)}}}
\def\gfr{{g_{(4)}}}
\def\SC{{S_{\rm C}}}
\def\RH{{R_{\rm H}}}

\def\wlBox{\mbox{
\raisebox{0.1cm}{$\widetilde{\mbox{\raisebox{-0.1cm}\fbox{\ }}}$}}}
\def\BBox{\mbox{\raisebox{0.1cm}\fbox{\ }}}
\def\htBox{\mbox{
\raisebox{0.1cm}{$\hat{\mbox{\raisebox{-0.1cm}{$\Box$}}}$}}}

\section{Introduction}

There are various ways to realize the Randall-Sundrum brane-world 
Universe \cite{RS}.
In particulary, having in mind the relation with AdS/CFT 
correspondence \cite{AdS} in refs.
\cite{HHR,NOZ} (for related discussion, see \cite{related}), 
the quantum creation of the brane-world thanks to  conformal 
anomaly of four-dimensional 
fields has been  discussed. The mechanism of 
refs. \cite{HHR,NOZ} has been applied to construct the 
(inflationary) Brane New World.

Recently much attention has been paid to dS/CFT correspondence 
\cite{strominger, dscft} which is similar in spirit to AdS/CFT 
correspondence. (For earlier proposals on dS/CFT duality, 
see \cite{desitter} and for recent related discussion of 
thermodynamics in de Sitter space, see \cite{thermodynamics}).
The reason why AdS/CFT can be expected is the isometry of 
$d+1$-dimensional anti-de Sitter space, which is $SO(d,2)$ 
symmetry. It is identical with the conformal symmetry of 
$d$-dimensional Minkowski space. We should note, however,  
$d+1$-dimensional de Sitter space has the isometry of 
$SO(d+1,1)$ symmetry, which can be a conformal symmetry of 
$d$-dimensional Euclidean space. Then it might be natural to 
expect the correspondence between $d+1$-dimensional de Sitter 
space and $d$-dimensional euclidean conformal symmetry (dS/CFT 
correspondence\cite{strominger, dscft}). 
In fact, the metric of 
$D=d+1$-dimensional anti de Sitter space (AdS) is given by
\be
\label{AdSm}
ds_{\rm AdS}^2=dr^2 + \e^{2r}\left(-dt^2 + \sum_{i=1}^{d-1}
\left(dx^i\right)^2\right)\ .
\ee
In the above expression, the boundary of AdS lies at 
$r=\infty$. If one exchanges the radial coordinate $r$ 
and the time coordinate $t$, we obtain the metric of the 
de Sitter space (dS): 
\be
\label{dSm}
ds_{\rm dS}^2=-dt^2 + \e^{2t}\sum_{i=1}^d
\left(dx^i\right)^2\ .
\ee
Here $x^d=r$. Then there is a boundary at $t=\infty$, where the 
Euclidean conformal field theory (CFT) can live and one expects 
dS/CFT correspondence as one more manifestation of holographic principle.
This may be very important as there are indications that our Universe has 
de Sitter phase in the past and in the future. Then, there appears very 
nice way to formulate some de Sitter gravitational physics in terms of 
the boundary QFT physics and vice-versa.

The purpose of the present paper is to 
 consider the possibility of  quantum creation of 
the inflationary brane in de Sitter bulk space in frames of mechanism of
refs. \cite{HHR,NOZ}. Note that such approach represents the generalization 
of so-called anomaly-driven inflation \cite{sta}. 
Moreover, the content of quantum fields on the brane may be chosen in such
a way, that it corresponds to euclidean CFT dual to 5d dS bulk space.
In this sense, one can understand that quantum creation
of dS brane-world occurs in frames of dS/CFT correspondence.
In \cite{NOZ} several cases corresponding to flat, sphere 
or hyperboloid (brane) embedded in 5d AdS space have been  considered.
 If we Wick-rotate 5-dimensional 
de Sitter space into the Euclidean signature, we obtain 5d 
sphere. Then one cannot embedd the 4d flat or hyperbolic brane in 
the bulk 5d space. That is why we consider only the case that  brane 
is 4d sphere. 

We will show that gravity on such de Sitter brane (despite the fact 
that bulk represents not AdS but dS space) may be localized 
using proposed dS/CFT correspondence. 
Then, our model may be understood as four-dimensional gravity coupled to 
some gauge theory. As a result the model turns out to some kind
 of trace-anomaly driven inflation,
which is known \cite{vilenkin} may become instable. As a result 
there is natural solution to end the inflationary phase.
The equations of moving brane in SdS background are also considered
and presented as FRW equations with quantum corrections.

\section{de Sitter brane-worlds}

The metric of  5 dimensional Euclidean de Sitter space 
that is 5d sphere is given by
\be
\label{dSi}
ds^2_{{\rm S}_5}=dy^2 + l^2 \sin^2 {y \over l}d\Omega^2_4\ .
\ee
Here $d\Omega^2_4$ describes the metric of ${\rm S}_4$ 
with unit radius. The coordinate $y$ is defined in 
$0\leq y \leq l\pi$. 
One also assumes the brane 
lies at $y=y_0$ 
and the bulk space is given by gluing two regions 
given by $0\leq y < y_0$. 

We start with the action $S$ which is the sum of 
the Einstein-Hilbert action $\SEH$ with positive 
cosmological constant, the Gibbons-Hawking 
surface term $\SGH$,  the surface counter term $S_1$\footnote{
The coefficient of $S_1$ cannot be determined from the condition of 
finiteness of the action as in AdS/CFT. However, using the 
renormailzation group method as in \cite{BVV}  this 
coefficient can be determined uniquely, see also  third paper 
in \cite{dscft}. 
} 
and the trace anomaly induced action 
${\cal W}$: 
\bea
\label{Stotal}
&& S=\SEH + \SGH + 2 S_1 + {\cal W}\ ,\quad 
\SEH={1 \over 16\pi G}\int d^5 x \sqrt{\gfv}\left(R_{(5)} 
 - {12 \over l^2}\right)\ , \nn
&& \SGH={1 \over 8\pi G}\int d^4 x \sqrt{\gfr}\nabla_\mu n^\mu 
\ ,\quad S_1= -{3 \over 8\pi Gl}\int d^4 x \sqrt{\gfr} \ ,\nn
&&{\cal W}= b \int d^4x \sqrt{\widetilde g}\widetilde F A 
 + b' \int d^4x \sqrt{\widetilde g}
\left\{A \left[2 {\wlBox}^2 
+\widetilde R_{\mu\nu}\widetilde\nabla_\mu\widetilde\nabla_\nu 
 - {4 \over 3}\widetilde R \wlBox^2 \right.\right. \nn
&& \left.\left. \qquad 
+ {2 \over 3}(\widetilde\nabla^\mu \widetilde R)\widetilde\nabla_\mu
\right]A 
+ \left(\widetilde G - {2 \over 3}\wlBox \widetilde R
\right)A \right\} \nn
&& \qquad -{1 \over 12}\left\{b''+ {2 \over 3}(b + b')\right\}
\int d^4x \sqrt{\widetilde g} \left[ \widetilde R - 6\wlBox A 
 - 6 (\widetilde\nabla_\mu A)(\widetilde \nabla^\mu A)
\right]^2 \ .
\eea 
Here the quantities in the  5 dimensional bulk spacetime are 
specified by the suffices $_{(5)}$ and those in the boundary 4 
dimensional spacetime  by $_{(4)}$. 
The factor $2$ in front of $S_1$ in (\ref{Stotal}) is coming from 
that we have two bulk regions which 
are connected with each other by the brane. 
In (\ref{Stotal}), $n^\mu$ is 
the unit vector normal to the boundary. In (\ref{Stotal}), 
one chooses the 4 dimensional boundary metric as 
\be
\label{tildeg}
\gfr_{\mu\nu}=\e^{2A}\tilde g_{\mu\nu}
\ee 
and we specify the 
quantities with $\tilde g_{\mu\nu}$ by using $\tilde{\ }$. 
$G$ ($\tilde G$) and $F$ ($\tilde F$) are the Gauss-Bonnet
invariant and the square of the Weyl tensor, which are given as
\footnote{We use the following curvature conventions:
\begin{eqnarray*}
&& R=g^{\mu\nu}R_{\mu\nu}\ ,\quad 
R_{\mu\nu}= R^\lambda_{\ \mu\lambda\nu} \\
&& R^\lambda_{\ \mu\rho\nu}=-\Gamma^\lambda_{\mu\rho,\nu}
+ \Gamma^\lambda_{\mu\nu,\rho}
- \Gamma^\eta_{\mu\rho}\Gamma^\lambda_{\nu\eta}
+ \Gamma^\eta_{\mu\nu}\Gamma^\lambda_{\rho\eta}\ ,\quad 
\Gamma^\eta_{\mu\lambda}={1 \over 2}g^{\eta\nu}\left(
g_{\mu\nu,\lambda} + g_{\lambda\nu,\mu} - g_{\mu\lambda,\nu} 
\right)\ .
\end{eqnarray*}}
\be
\label{GF}
G=R^2 -4 R_{ij}R^{ij}
+ R_{ijkl}R^{ijkl} \ ,\quad 
F={1 \over 3}R^2 -2 R_{ij}R^{ij}
+ R_{ijkl}R^{ijkl} \ ,
\ee
In the effective action (\ref{actions2}) induced by brane quantum 
matter, in general, with $N$ real scalar, $N_{1/2}$ 
Dirac spinor, $N_1$ vector 
fields, $N_2$  ($=0$ or $1$) gravitons and $N_{\rm HD}$ higher 
derivative conformal scalars, $b$, $b'$ and $b''$ are
\bea
\label{bs}
&& b={N +6N_{1/2}+12N_1 + 611 N_2 - 8N_{\rm HD} 
\over 120(4\pi)^2}\ ,\nn 
&& b'=-{N+11N_{1/2}+62N_1 + 1411 N_2 -28 N_{\rm HD} 
\over 360(4\pi)^2}\ ,\quad b''=0\ .
\eea
Usually, $b''$ may be changed by the finite renormalization 
of local counterterm in the gravitational effective action 
but as we will see later, the term proportional 
to $\left\{b''+ {2 \over 3}(b + b')\right\}$ in (\ref{actions2}), 
and therefore $b''$ does not contribute to the equations 
describing the nucleation of the brane.
Nevetheless, this parameter plays an important role in tensor perturbations,
what leads to decay of de Sitter space (end of inflation).

For typical examples motivated by AdS/CFT (and presumbly by dS/CFT 
because central charges are the same in AdS/CFT or dS/CFT)
correspondence 
one has:
a) ${\cal N}=4$ $SU(N)$ SYM theory 
$b=-b'={N^2 -1 \over 4(4\pi )^2}$, 
b) ${\cal N}=2$ $Sp(N)$ theory 
$b={12 N^2 + 18 N -2 \over 24(4\pi)^2}$, 
$b'=-{12 N^2 + 12 N -1 \over 24(4\pi)^2}$. 
 Note that $b'$ is negative in the above cases.

We should also note that ${\cal W}$ in (\ref{actions2}) is defined up to 
conformally invariant functional, which cannot be determined 
from only the conformal anomaly. The conformally flat space is a pleasant 
exclusion where anomaly induced effective action is defined uniquely.
However, one can argue that such conformally invariant functional gives 
next to leading contribution as mass parameter of regularization 
may be adjusted to be arbitrary small (or large).

The metric of ${\rm S}_4$ with the unit radius is given by
\be
\label{S4metric1}
d\Omega^2_4= d \chi^2 + \sin^2 \chi d\Omega^2_3\ .
\ee
Here $d\Omega^2_3$ is described by the metric of 3 dimensional 
unit sphere. If one changes the coordinate $\chi$ to 
$\sigma$ by $\sin\chi = \pm {1 \over \cosh \sigma}$, 
one obtains\footnote{
If we Wick-rotate the metric by $\sigma\rightarrow it$, we 
obtain the metric of de Sitter space:
\[
d\Omega^2_4\rightarrow ds_{\rm dS}^2
= {1 \over \cos^2 t}\left(-dt^2 + d\Omega^2_3\right)\ .
\]
}
\be
\label{S4metric2}
d\Omega^2_4= {1 \over \cosh^2 \sigma}\left(d \sigma^2 
+ d\Omega^2_3\right)\ .
\ee
Then one assumes 
the metric of 5 dimensional space time as follows:
\be
\label{metric1}
ds^2=dy^2 + \e^{2A(y,\sigma)}\tilde g_{\mu\nu}dx^\mu dx^\nu\ ,
\quad \tilde g_{\mu\nu}dx^\mu dx^\nu\equiv l^2\left(d \sigma^2 
+ d\Omega^2_3\right)
\ee
and one identifies $A$ and $\tilde g$ in (\ref{metric1}) with those in 
(\ref{tildeg}). Then $\tilde F=\tilde G=0$, 
$\tilde R={6 \over l^2}$ etc. 
Due to the assumption (\ref{metric1}), the actions in (\ref{Stotal}) 
have the following forms:
\bea
\label{actions2}
&& \SEH= {l^4 V_3 \over 16\pi G}\int dy d\sigma \left\{\left( -8 
\partial_y^2 A - 20 (\partial_y A)^2\right)\e^{4A} \right. \nn
&& \qquad \left. +\left(-6\partial_\sigma^2 A 
 - 6 (\partial_\sigma A)^2 
+ 6 \right)\e^{2A} - {12 \over l^2} \e^{4A}\right\} \nn
&& \SGH= {l^4 V_3 \over 2\pi G}\int d\sigma \e^{4A} 
\partial_y A \ ,\quad 
S_1= - {3l^3 V_3 \over 8\pi G}\int d\sigma \e^{4A} \nn
&& {\cal W}= V_3 \int d\sigma \left[b'A\left(2\partial_\sigma^4 A
 - 8 \partial_\sigma^2 A \right) 
 - 2(b + b')\left(1 - \partial_\sigma^2 A 
 - (\partial_\sigma A)^2 \right)^2 \right]\ .
\eea
Here $V_3$ is the volume or area of the unit 3 sphere. 

In the bulk, one obtains the following equation of motion 
from $\SEH$ by the variation over $A$:
\be
\label{eq1}
0= \left(-24 \partial_y^2 A - 48 (\partial_y A)^2 
 - {48 \over l^2}
\right)\e^{4A} + {1 \over l^2}\left(-12 \partial_\sigma^2 A 
- 12 (\partial_\sigma A)^2 + 12\right)\e^{2A}\ ,
\ee
which corresponds to one of the Einstein equations. 
Then one finds solutions, $A_S$, which correspond to 
the metric  dS$_5$ in (\ref{dSi}) with (\ref{S4metric2}). 
\be
\label{blksl}
A=A_S=\ln\sin{y \over l} - \ln \cosh\sigma\ .
\ee

On the brane at the boundary, 
one gets the following equation:
\bea
\label{eq2}
0&=&{48 l^4 \over 16\pi G}\left(\partial_y A - {1 \over l}
\right)\e^{4A}
+b'\left(4\partial_\sigma^4 A - 16 \partial_\sigma^2 A\right) \nn
&& - 4(b+b')\left(\partial_\sigma^4 A + 2 \partial_\sigma^2 A 
 - 6 (\partial_\sigma A)^2\partial_\sigma^2 A \right)\ .
\eea
We should note that the contributions from $\SEH$ and $\SGH$ are 
twice from the naive values since we have two bulk regions which 
are connected with each other by the brane. 
Substituting the bulk solution $A=A_S$ in (\ref{blksl}) into 
(\ref{eq2}) and defining the radius $R$ of the brane by
$R\equiv l\sin{y_0 \over l}$, one obtains
\be
\label{slbr2}
0={1 \over \pi G}\left({1 \over R}\sqrt{1 - {R^2 \over l^2}}
 - {1 \over l}\right)R^4 + 8b'\ .
\ee
One sees that eq.(\ref{slbr2}) does not depend on $b$. 
First we should note $0\leq R\leq l$ by  definition. 
Even in the case that there is no quantum contribution 
from the matter on the brane, that is, $b'=0$, Eq.(\ref{slbr2}) 
has a solution:
\be
\label{Csol}
R^2=R_0^2\equiv 
{l^2 \over 2}\ \mbox{or}\ {y_0 \over l}={\pi \over 4}, 
{3\pi \over 4}\ .
\ee
In Eq.(\ref{slbr2}), the first term 
${R^3 \over \pi G}\sqrt{1 - {R^2 \over l^2}}$ 
corresponds to the gravity, which makes the radius $R$ larger. 
On the other hand, the second term 
$-{R^4 \over \pi Gl}$ corresponds to the tension, which makes 
$R$ smaller. When $R<R_0$,  gravity becomes larger than the 
tension and when $R>R_0$, vice versa. Then both of the solutions in 
(\ref{Csol}) are stable. Although it is not clear from 
(\ref{slbr2}), $R=l$ (${y \over l}={\pi \over 2}$) corresponds 
to the local maximum. 
Hence, the possibility of creation of inflationary brane in 
de Sitter bulk 
is possible already on classical level, even in situation when brane 
tension is fixed by holographic RG. That is qualitatively different from 
the case of AdS bulk where only quantum effects led to creation of
inflationary 
brane \cite{NOZ,HHR} (when brane tension was not free 
parameter).

Let us make several remarks about properties of dS brane-world.
There is an excellent explanation \cite{HHR}  why gravity 
is trapped on the brane in the AdS spacetime. This uses AdS$_5$/CFT$_4$ 
correspondence  and the surface counter terms. This can be generalized to 
the brane in dS spacetime by using proposed dS/CFT correspondence. 

In \cite{BBM} it has been shown that  the bulk action diverges
in de Sitter space when we substitute the classical 
solution, which is the fluctuation around the de Sitter space 
in (\ref{dSm}). In other words,  counterterms are necessary again. The 
divergence 
occurs since the volume of the space diverges when $t\rightarrow 
\infty$ (or $t\rightarrow -\infty$ after replacing $t$ by $-t$ 
in another patch). Then we should put the counterterms on the 
space-like branes which lie at $t\rightarrow \pm\infty$. 
Therefore dS/CFT correspondence should be given by
\bea
\label{lc1}
&& \e^{-W_{\rm CFT}}=\int [dg][d\varphi]\e^{-S_{\rm dS\,grav}}\ , 
\quad S_{\rm dS\,grav}=\SEH + \SGH + S_1 + S_2 + \cdots, \nn 
&& \SEH={1 \over 16\pi G}\int d^5 x \sqrt{-\gfv}\left(R_{(5)} 
 - {12 \over l^2} + \cdots \right)\ , \nn 
&& \SGH={1 \over 8\pi G}\int_{M_4^+ + M_4^-} d^4 x 
\sqrt{\gfr}\nabla_\mu n^\mu, \\
&& S_1= {3 \over 8\pi G l}\int_{M_4^+ + M_4^-} d^4 x 
\sqrt{\gfr}\ , \quad 
S_2= {l \over 32\pi G }\int_{M_4^+ + M_4^-} d^4 x 
\sqrt{\gfr}\left(R_{(4)} + \cdots \right)\ ,  \cdots \ . 
\nonumber
\eea
Here $S_1$, $S_2$, $\cdots$ correspond to the surface counter terms, 
which cancell the divergences in the bulk action and $M_4^\pm$ 
expresses the boundary at $t\rightarrow \pm \infty$.  

Let us consider two copies of the de Sitter spaces dS$_{(1)}$ 
and dS$_{(2)}$. We also put one or two of the space-like 
branes, which can be 
regarded as boundaries  connecting the two bulk de Sitter 
spaces, at finite $t$. Then if one takes the 
following action $S$ instead of $S_{\rm dS\,grav}$, 
\be
\label{lc2}
S=\SEH + \SGH + 2S_1=S_{\rm dS\,grav} + S_1 - S_2 - \cdots,
\ee
we obtain the following boundary theory in terms of 
the partition function:
\bea
\label{lc3}
&& \int_{{\rm dS_5^{(1)} + dS_5^{(1)}} +M_4^+ + M_4^-} 
[dg][d\varphi]\e^{-S} 
= \left(\int_{{\rm dS}_5} [dg][d\varphi]\e^{-\SEH - \SGH - S_1} 
\right)^2 \nn
&& \quad =\e^{2S_2 + \cdots}\left(\int_{{\rm dS}_5} [dg][d\varphi]
\e^{-S_{\rm grav}} \right)^2 =\e^{-2W_{\rm CFT}+2S_2 + \cdots}\ .
\eea
Since $S_2$ can be regarded as the Einstein-Hilbert action on 
the boundary, the gravity on the boundary becomes dynamical. 
In other words, there is strong indication that our brane-world model at 
some conditions may be effectively described by 4d gravity interacting with 
some gauge theory.

Now we consider the quantum effects ($b'\neq 0$ case) on 
the brane in (\ref{slbr2}). Let us define a function 
$F(R^2)$ as follows:
\be
\label{F}
F(R^2)={1 \over \pi G}\left({1 \over R}\sqrt{1 - {R^2 \over l^2}}
 - {1 \over l}\right)R^4 \ .
\ee
Then one can easily find
\bea
\label{prF}
&& F(0)=F\left({l^2 \over 2}\right)=0 \ ,\quad 
F(l^2)=-{l^3 \over \pi G} \ ,\nn
&& F(R^2)\begin{array}{ll}
 >0 \quad & \mbox{when}\ 0<R^2<{l^2 \over 2} \\
 <0 \quad & \mbox{when}\ {l^2 \over 2}<R^2 \leq l^2 \\
\end{array} \ .
\eea
The function $F(R^2)$ has a maximum 
\be
\label{mF0}
F=F_m\equiv {l^3 \over 16\pi G}\left(-26 
+ 35 \sqrt{1-{9 \over 50}}\right)
\ee
when 
\be
\label{mF}
R^2=R_m^2\equiv {5l^2 \over 4}\left(1-\sqrt{1-{9 \over 50}}
\right)<{l^2 \over 2}\ .
\ee
The above results tell 
\begin{enumerate}
\item When $-8b'> F_m$ or $-8b' < -{l^3 \over \pi G}$, there 
is no solution in Eq.(\ref{slbr2}). That is, the quantum effect 
exhibits the creation of the inflationary brane world. 
\item When $0<-8b'< F_m$, there appear two solutions in 
(\ref{slbr2}). The solution with larger radius $R$ corresponds to 
the classical one in (\ref{Csol}) but the radius $R$ in the 
solution is smaller then that in the classical one. 
In other words, quantum effects try to prevent inflation. The solution 
with smaller radius can be regarded as the solution  generated 
by only quantum effects on the brane. Anyway the radii $R$ in 
both of the solutions are smaller than that in the classical one 
(\ref{Csol}). Since ${1 \over R}$ corresponds to the rate of the 
expansion of the universe when S$_4$ is Wick-rotated into 4d de 
Sitter space, the quantum effect makes the rate larger. 
\item When $0>-8b'>-{l^3 \over \pi G}$, which is rather exotic 
case since usualy $b'$ is negative as in (\ref{bs}), 
Eq.(\ref{slbr2}) has unique solution corresponding to the 
solution in the classical case (\ref{Csol}) and the quantum 
effect on the brane makes the radius $R$ larger. 
\end{enumerate}

The de Sitter brane may be thought as inflationary brane.
The natural question then appears how such inflation may become instable?
The answer goes in the same way as in anomaly-driven inflation \cite{HHR2}.
Despite the fact that the term in the effective action related with 
coefficient $b''$ does not give contribution to the equations of the motion,
it is important in the study of perturbations. It may be shown, by analogy 
with \cite{HHR2, vilenkin} that there is some bound for this coefficient 
which makes the inflation to be instable. Indeed, we got the alternative 
description of the brane-world as some 4d gravity 
with matter. Then, the analysis of instability of inflation may be 
repeated in all details and values of parameter $b''$ which ensure the
instability may be found. We will not go to the details of such analysis 
as it repeats very much the same done recently in \cite{HHR2}.
So principal possibility of the end of brane inflation exists.

If we Wick-rotate S$_4$ into the Lorentzian signature, we can 
obtain 4d de Sitter space. In some choice of the time 
coordinate, the de Sitter space can be regarded as an inflationary 
universe. The rate of the inflation corresponds to the inverse 
of the radius of S$_4$. 
Hence, we estimated the role of quantum effects to creation of 
de Sitter brane-world. As one can see the brane 
inflation occurs on classical as well as on quantum levels in 
5d de Sitter bulk space. 
Quantum effects not only decrease the 
radius of classically created de Sitter brane but also can 
produce another (purely quantum) de Sitter brane.

Let us give some remarks  about the Wick-rotation of the above 
obtained brane solution. There are several ways for the 
Wick-rotation of the sphere into de Sitter space. 
The metric of S$_5$ can be expressed as
\be
\label{S5a}
ds_{{\rm S}_5}^2=l^2\left(d\chi^2 + \sin^2\chi \left(d\eta^2 
+ \sin^2d\Omega_2^2\right)\right)\ .
\ee
Here $d\Omega_2^2$ is the metric of 2 dimensional sphere. 
The brane  S$_4$ can be embedded into S$_5$ by putting the 
coordinate $\chi$ to be a constant: $\chi=\chi_0$. Then the 
metric of S$_4$ has the folowing form
\be
\label{S4a}
ds_{{\rm S}_4}^2=l^2 \sin^2\chi_0 \left(d\eta^2 
+ \sin^2d\Omega_2^2\right)\ .
\ee
If we further write $d\Omega_2^2$ as 
\be
\label{S2a}
d\Omega_2^2=d\theta^2 + \sin^2\theta d\phi^2 
\ee
and Wick-rotate $\phi$ by 
\be
\label{phi}
\phi\rightarrow it_1\ ,
\ee
we obtain the static 4d de Sitter brane in static 5d de 
Sitter bulk space:
\bea
\label{W1}
ds_{{\rm S}_5}^2&\rightarrow& l^2\left(d\chi^2 
+ \sin^2\chi \left(d\eta^2 + \sin^2\eta
\left(d\theta^2 - \sin^2\theta dt_1^2\right)\right)\right) \ ,\nn
ds_{{\rm S}_4}^2&\rightarrow& l^2 \sin^2\chi_0 
\left(d\eta^2 + \sin^2\eta\left(d\theta^2 - \sin^2\theta dt_1^2
\right)\right) \ .
\eea
On the other hand, if we Wick-rotate the coordinate $\chi$ by
\be
\label{W2}
\chi\rightarrow {\pi \over 2}+it_2\ ,\quad 
\chi_0\rightarrow {\pi \over 2}+it_0\ ,
\ee
the brane becomes the space-like surface of S$_4$ in  5d 
de Sitter space, which can be regarded as the inflationary 
universe 
\bea
\label{W3}
ds_{{\rm S}_5}^2&\rightarrow&l^2\left(-dt_2^2 + \cosh^2 t_2 
\left(d\eta^2 + \sin^2\eta\Omega_2^2\right)\right)\ ,\nn
ds_{{\rm S}_4}^2&\rightarrow&l^2 \cosh^2 t_0 
\left(d\eta^2 + \sin^2\eta\Omega_2^2\right)\ .
\eea
When we Wick-rotate the coordinate $\eta$ by 
\be
\label{W4}
\eta\rightarrow {\pi \over 2}+it_3\ ,
\ee
we obtain 4d de Sitter brane, which can be regarded as the 
inflationary universe
\bea
\label{W5}
ds_{{\rm S}_5}^2&\rightarrow&l^2\left(-\sin^2\chi dt_3^2 
+ d\chi^2 + \sin^2\chi \cosh^2 t_3 d\Omega_3^2\right)\ ,\nn
ds_{{\rm S}_4}^2&\rightarrow&l^2 \sin^2\chi_0 
\left(- dt_3^2 + \cosh^2 t_3 d\Omega_3^2\right)\ .
\eea
Here $d\Omega_3^2$ is the metric of the 3d unit sphere. 
The expression for 5d de Sitter bulk space is not so 
conventional. 

In \cite{HHR2}, trace anomaly driven inflationary 
model \cite{sta} has been studied  using AdS/CFT 
correspondence. The brane, which is Euclidean 4 sphere, 
can be nucleated due to quantum effects in AdS spacetime 
and it can be regarded as an instanton \cite{HHR,NOZ}. 
The brane can be analytically continued into 
the Lorentzian signature and the de Sitter space is nucleated. 
The 4d de Sitter space can be identified with the inflationary 
universe \cite{sta}. 
In \cite{HHR2}, as in the original model in \cite{sta}, it was 
shown that the de Sitter space is instable and  decays into 
the matter dominant FRW universe. For such a decay, the term 
with coefficient $b''$ of ${\cal W}$ in (\ref{Stotal}) is 
important and affect the perturbation of the tensor part in the 
metric. Such term also appears as $\alpha'$ 
corrections in the string theory. Therefore this term maybe induced 
by the square of the scalar curvature  even if on quantum level we took 
$b''=0$. In \cite{HHR2},  following 
the arguments from \cite{vilenkin}, the inflation occurs until
\be
\label{inf1}
T=t_*\sim -2\alpha(\gamma - 1)R\ .  
\ee
Here we can shoose $\alpha = {16\pi^2 \over N^2}b''$ when we 
consider ${\cal N}=4$ $SU(N)$ Yang-Mills theory on the brane.
The parameter $\gamma$ is related with the initial perturbation 
from the de Sitter solution and can be expressed as the 
perturbation of the Hubble parameter $H$ when the de Sitter 
universe is nucleated:
\be
\label{inf2}
\gamma={1 \over 2}\ln \left({2H_0 \over H_0 - H}\right)\ .
\ee
Here $H_0$ is the inverse of $R$ in (\ref{slbr2}), which is 
the radius of the 4d sphere. 
Thus, the arguments are presented which show
 the possibility 
to end the inflation.

\section{FRW brane in 5d SdS black hole}

We now consider the situation that radius depends 
on the ``time'' coordinate. Taking 5d Schwarzschild-
de Sitter (SdS) black hole background, the obtained brane equation, which 
describes the dynamics of the brane, can be regarded as the induced 
Friedmann-Robertson-Walker (FRW) equation. 

Starting with the Minkowski signature action one gets 
 the following 
equation which generalizes the 
classical brane equation (\ref{eq2}):
\bea
\label{eq2bb}
0&=&{48 l^4 \over 16\pi G}\left(A_{,z} 
 - {1 \over l}\right)\e^{4A}
+b'\left(4 \partial_\tau^4 A + 16 \partial_\tau^2 A\right) \nn
&& - 4(b+b')\left(\partial_\tau^4 A - 2 \partial_\tau^2 A 
 - 6 (\partial_\tau A)^2\partial_\tau^2 A \right) \ .
\eea
In (\ref{eq2bb}), one 
uses the form of the metric as 
\be
\label{metric1b}
ds^2=dz^2 + \e^{2A(z,\tau)}\tilde g_{\mu\nu}dx^\mu dx^\nu\ ,
\quad \tilde g_{\mu\nu}dx^\mu dx^\nu\equiv l^2\left(-d \tau^2 
+ d\Omega^2_3\right)\ .
\ee
Here $d\Omega^2_3$ corresponds to the metric of 3 dimensional 
unit sphere. 
As a bulk space,  one takes 5d Schwarzschild-
de Sitter spacetime, whose metric is given by
\be
\label{dSS} 
ds_{\rm dS-S}^2 = {1 \over h(a)}da^2 - h(a)dt^2 
+ a^2 d\Omega_3^2 \ ,\ \ 
h(a)= -{a^2 \over l^2} + 1 - {16\pi GM \over 3 V_3 a^2}\ .
\ee
Here $V_3$ is the volume of the unit 3 sphere.  
If one chooses new coordinates $(z,\tau)$ by
\be
\label{cc1}
{\e^{2A} \over h(a)}A_{,z}^2 - h(a) t_{,z}^2 = 1 \ ,
\quad {\e^{2A} \over h(a)}A_{,z}A_{,\tau} - h(a)t_{,z} t_{,\tau}
= 0 \ ,\quad 
{\e^{2A} \over h(a)}A_{,\tau}^2 - h(a) t_{,\tau}^2 
= -\e^{2A}\ .
\ee
the metric takes the form (\ref{metric1b}). Here $a=l\e^A$.
Furthermore choosing a coordinate $\tilde t$ by 
$d\tilde t = l\e^A d\tau$, 
the metric on the brane takes FRW form: 
\be
\label{e3}
\e^{2A}\tilde g_{\mu\nu}dx^\mu dx^\nu= -d \tilde t^2  
+ l^2\e^{2A} d\Omega^2_3\ .
\ee
Solving Eqs.(\ref{cc1}), one gets
\be
\label{e4}
H^2 = A_{,z}^2 - h\e^{-2A}= A_{,z}^2 + {1 \over l^2}
 - {1 \over a^2} + {16\pi GM \over 3 V_3 a^4}\ .
\ee
Here the Hubble constant $H$ is defined by
$H={dA \over d\tilde t}$. Then  using the brane equation 
(\ref{eq2bb}), we obtain 
\bea
\label{e10}
H^2 &=& - {1 \over a^2} 
+ {8\pi G_4 \rho \over 3} \\
\label{e10b}
\rho&=&{l \over a}\left[ {M \over V_3 a^3} \right. 
+ {3a \over 16\pi G}\left[
\left[{1 \over l} + {\pi G \over 3}\left\{ 
-4b'\left(\left(H_{,\tilde t \tilde t \tilde t} + 4 H_{,\tilde t}^2 
+ 7 H H_{,\tilde t\tilde t} \right.\right.\right.\right. \right.\nn
&& \left.\left. + 18 H^2 H_{,\tilde t} + 6 H^4\right) 
+ {4 \over a^2} \left(H_{,\tilde t} + H^2\right)\right) 
+ 4(b+b') \left(\left(H_{,\tilde t \tilde t \tilde t} 
+ 4 H_{,\tilde t}^2 \right. \right. \nn
&& \left.\left.\left.\left.\left.\left. + 7 H H_{,\tilde t\tilde t} 
+ 12 H^2 H_{,\tilde t} \right) - {2 \over a^2} 
\left(H_{,\tilde t} + H^2\right)\right) \right\}\right]^2
 + {1 \over l^2} \right]\right]\ .
\eea
This can be regarded as the quantum FRW equation of the brane 
universe. It again admits  quantum-corrected dS brane solutions. Here 4d
Newton
constant $G_4$ is given by
\be
\label{e12}
G_4={2G \over l}\ .
\ee

Note that forgetting about quantum corrections we have just 
standard FRW equation with some energy density $\rho$ 
expressed in terms of 5d parameters:
\be
\label{ee1}
\rho_c={Ml \over V_3 a^4} + {3 \over 8\pi G l}\ ,
\ee
where, from the point of view of 4d spacetime, 
the first term can be regarded as the contribution from 
the radiation and the second term as that from the cosmological 
constant. Since the energy density $\rho$ in (\ref{e10}) 
contains the higher derivative terms, the quantum correction 
 becomes important when the size of the universe changes 
rapidly, as in the early stage of the universe.
 
In a sense, brane FRW approach represents the attempt to describe 
the 4d cosmology in terms of the observer who knows about 
our 5d brane-world. As we already showed in the specific case 
of the previous section, these FRW equations may lead to 
reasonable early time cosmological behaviour.

Note that matter content may be chosen in such a way that brane CFT is dual
to one of bulk SdS BH backgrounds in accord with dS/CFT correspondence.
Further by differentiating Eq.(\ref{e10}) with respect to 
$\tilde t$, one obtains the second FRW equation
\bea
\label{e11}
H_{,\tilde t} &=&  {1 \over a^2} - 4\pi G_4(\rho + p) \\
\label{e11b}
\rho + p &=& {l \over a}\left[
 {4 M \over 3 V_3 a^3} \right. - {1 \over 24l^3 H}\left[{1 \over l} 
+ {\pi G \over 3}\left\{ 
-4b'\left(\left(H_{,\tilde t \tilde t \tilde t} + 4 H_{,\tilde t}^2 
+ 7 H H_{,\tilde t\tilde t} \right.\right.\right.\right. \nn
&& \left.\left. + 18 H^2 H_{,\tilde t} + 6 H^4\right) 
+ {4 \over a^2} \left(H_{,\tilde t} + H^2
\right)\right) \nn
&& + 4(b+b') \left(\left(H_{,\tilde t \tilde t \tilde t} 
+ 4 H_{,\tilde t}^2 + 7 H H_{,\tilde t\tilde t} 
+ 12 H^2 H_{,\tilde t} \right) 
\left.\left. - {2 \over a^2} \left(H_{,\tilde t} + H^2\right) 
\right)\right\}\right] \nn
&& \times \left\{ 
-4b'\left(\left(H_{,\tilde t \tilde t \tilde t \tilde t} 
+ 15 H_{,\tilde t} H_{\tilde t\tilde t} 
+ 7 H H_{,\tilde t\tilde t\tilde t} 
+ 18 H^2 H_{,\tilde t\tilde t} 
+ 36 H H_{,\tilde t}^2 \right.\right.\right. \nn
&& \left.\left. + 24 H^3 H_{,\tilde t} \right)
+ {4 \over a^2} \left(H_{,\tilde t\tilde t} - 2 H^3\right) \right) 
+ 4(b+b') \left(\left(H_{,\tilde t \tilde t \tilde t \tilde t} 
+ 15 H_{,\tilde t} H_{,\tilde t\tilde t} \right.\right. \nn
&& \left.\left. 
+ 7 H H_{,\tilde t\tilde t\tilde t} + 12 H^2 H_{,\tilde t\tilde t}
+ 24 H H_{,\tilde t}^2 \right)
\left.\left. - {2 \over a^2} \left(H_{,\tilde t\tilde t} 
 - 2H^2\right)\right) \right\}\right]\ .
\eea
The quantum corrections from CFT are included into the definition of
energy (pressure). These quantum corrected FRW equations are written
from quantum-induced brane-world perspective. 

The Schwarzschild- de Sitter black hole solution in (\ref{dSS}) 
has a horizon at $a=a_H$, where $h(a)$ vanisif the higher 
derivative of the Hubble constant $H$ is 
large, the quantum correction becomes essential. 

The Schwarzschild- de Sitter black hole solution in (\ref{dSS}) 
has a horizon at $a=a_H$, where $h(a)$ vanishes:
\be
\label{eq16}
h(a_H)= -{a_H^2 \over l^2} + 1 - {16\pi GM \over 3 V_3 a_H^2}=0\ .
\ee
The solutions of Eq.(\ref{eq16}) are given by
\be
\label{H1}
a_{H\pm}^2={l^2\pm \sqrt{l^4 - 4\mu l^2} \over 2}\ ,
\quad \mu \equiv {16\pi G M \over 3V_3}
\ee
if 
\be
\label{H1b}
4\mu \leq l^2\ .
\ee
The cosmological horizon lies at $a=a_{H+}$ and the black 
hole one at $a=a_{H-}$. When 
\be
\label{H2}
4\mu=l^2\ ,
\ee
we have the extremal solution or Nariai space, where 
the horizons coincide with each other. 

We now consider the case that the brane is static. 
Then since $H=0$, the FRW equation (\ref{e10}) has the 
following form:
\be
\label{BB1}
0={2 \over l^2} - {1 \over a^2} + {\mu \over a^4}\ ,
\ee
which has solutions 
\be
\label{BB2}
a_{0\pm}={l^2 \pm \sqrt{l^4 - 8\mu l^2} \over 4}
\ee
if 
\be
\label{BB3}
8\mu \leq l^2\ .
\ee
Since 
\be
\label{BB4}
a_{H-}^2<a_{0\pm}^2<a_{H+}^2\ ,
\ee
the brane can exist between the black hole horizon and the 
cosmological one. In (\ref{BB2}), $a_{0+}$ corresponds to the 
solution in (\ref{Csol}) since the solution does not vanish even 
if $\mu=0$. This tells the solution corresponding to $a_{0+}$ is 
stable. Then the solution corresponding to $a_{0-}$ should be 
unstable.

The brane FRW like equations in (\ref{e10},\ref{e11}) 
are rather different from the corresponding equations obtained 
in this frame in AdS/CFT correspondence \cite{SV} (see also \cite{nooqr}).
In situation without  
 quantum corrections on the brane one gets 
\bea
\label{e10cl}
&& H^2 = - {1 \over a^2} 
+ {8\pi G_4 \rho \over 3} \ ,\quad 
\rho={l \over a}\left[ {M \over V_3 a^3} 
+ {3a \over 8\pi Gl^2}\right]\ ,\\
\label{e11cl}
&& H_{,\tilde t} =  {1 \over a^2} - 4\pi G_4(\rho + p) \ ,\quad
\rho + p = {l \over a}\cdot {4 M \over 3 V_3 a^3} \ .
\eea
In \cite{SV}, the second term in $\rho$ did not appear.
Furthermore, we have 
\be
\label
-\rho+3p=- {3 \over 2\pi Gl}\neq 0\ ,
\ee
which tells that the matter on the brane would not be 
conformal. Then the relation of cosmological entropy with Cardy formula
\cite{Cardy} 
is not very clear. The difference $\lambda$ of $\rho$ from 
the AdS/CFT case is a constant 
\be
\label{lambda}
\lambda=\rho - {l \over a}\cdot {M \over V_3 a^3} = 
{3 \over 8\pi Gl} \ .
\ee
which indicates that the effective cosmological constant 
on the brane does not vanish. It would be really interesting 
to investigate this question in order to understand if the possibility
to obtain the cosmological entropy bounds (so-called Cardy-Verlinde formula 
\cite{others}) exists in the present (de Sitter brane-world) context.

\section{Discussion}

In summary, we discussed two 5d de Sitter bulk spaces (in different
coordinate systems) connected by 4d de Sitter boundary playing 
the role of inflationary Universe. It is demonstrated that even 
in situation when brane tension is fixed by holographic 
RG the possibility of creation of such brane-world is not zero,
the radius of de Sitter brane may be defined (unlike to the case of 
two 5d AdS bulks \cite{HHR,NOZ}). Taking into account quantum 
brane fields via corresponding trace anomaly-induced effective action 
we proved the possibility of quantum creation of de Sitter brane-world.
The role of quantum effects is to decrease classical de Sitter brane 
radius, as well as to induce purely quantum de Sitter brane with 
even smaller radius. 
It is important to note that brane CFT may be chosen to be dual 
to one of de Sitter bulk spaces which may be relevant for relation
of brane-world approach with dS/CFT correspondence.
This dS/CFT correspondence plays the important role in the demonstration
that 4d gravity is localized and hence 4d inflationary brane may be
described as some variant of anomaly-driven inflation (with the possibility
to end the inflation).
Finally, we considered moving (time-dependent) de Sitter brane in
5d SdS BH when quantum brane fields again contribute to effective 
action. The quantum creation of 4d de Sitter Universe is again 
possible. 
Using special system of coordinates where brane equations look like 
FRW equations the comparison of such (quantum corrected) 
FRW-like equations in SdS BH bulk with the ones in SAdS BH  is done.


\begin{thebibliography}{99}
\bibitem{RS}L. Randall and R. Sundrum,
 {\sl Phys.Rev.Lett.} {\bf 83} (1999) 3370, hep-th/9905221;
 {\sl Phys.Rev.Lett.}  {\bf 83} (1999) 4690, hep-th/9906064.
\bibitem{AdS}  J.M. Maldacena, 
{\sl Adv.Theor.Math.Phys.} {\bf 2} (1998) 231;
E. Witten, {\sl Adv.Theor.Math.Phys.} {\bf 2} (1998) 253;
S. Gubser, I. Klebanov and A. Polyakov, {\sl Phys.Lett.} 
{\bf B428} (1998) 105;
O. Aharony, S. Gubser, J. Maldacena, H. Ooguri and Y. Oz,
{\sl Phys.Repts.} {\bf 323} (2000) 183.
\bibitem{HHR} S.W. Hawking, T. Hertog and H.S. Reall,
{\sl Phys.Rev.} {\bf D62} (2000) 043501, hep-th/0003052.
\bibitem{NOZ} S. Nojiri, S.D. Odintsov and S. Zerbini,
{\sl Phys.Rev.} {\bf D62} (2000) 064006, hep-th/0001192; 
 S. Nojiri and S.D. Odintsov, 
{\sl Phys.Lett.} {\bf B484} (2000) 119, 
hep-th/0004097; hep-th/0105160.
\bibitem{related} 
S. Nojiri, O. Obregon and S.D. Odintsov, 
{\sl Phys.Rev.} {\bf D62} (2000) 104003, hep-th/0005127;
L. Anchordoqui, C. Nunez and K. Olsen, hep-th/0007064;
K. Koyama and J. Soda, hep-th/0101164;
S. Nojiri, O. Obregon, S.D. Odintsov and V.I. Tkach, 
{\sl Phys.Rev.} {\bf D64} (2001) 043505, 
hep-th/0101003;
T. Shiromizu and D. Ida, hep-th/0102035;
M. Perez-Victoria, hep-th/0105048;
L. Anchordoqui, J. Edelstein, C. Nunez, S. Bergliaffa, M. Schvellinger, 
M. Trobo and F.Zyserman, hep-th/0106127.
\bibitem{strominger} A. Strominger, hep-th/0106113;
\bibitem{desitter}C.M. Hull, hep-th/9806146, {\sl JHEP} {\bf 9807} (1998)
021; E. Witten, hep-th/0106109;
A. Volovich, hep-th/0101176;
V. Balasubramanian, P. Horava and D. Minic, hep-th/0103171.
\bibitem{thermodynamics}
J. Maldacena and A. Strominger, {\sl JHEP} {\bf 9802} (1998) 014;
M. Banados, T. Brotz and M.E. Ortiz, {\sl Phys.Rev.} {\bf D59} (1999)
 046002; W.T. Kim, {\sl Phys.Rev.} {\bf D59} (1999) 047503;
F. Lin and Y. Wu, {\sl Phys.Lett.} {\bf B453} (1999) 028;
C.M. Hull and R. Khury, {\sl Nucl.Phys.} {\bf B575} (2000) 231;
S. Hawking, J. Maldacena and A. Strominger, {\sl JHEP} {\bf 0105}
(2001) 001; R. Bousso, {\sl JHEP} {\bf 0104} (2001) 035;
T. Banks and M. Fischler, hep-th/0102077;
\bibitem{dscft}
P.O. Mazur and E. Mottola, hep-th/0106151;
%
M. Li, hep-th/0106184;
%
S. Nojiri and S.D. Odintsov, {\sl Phys.Lett.} {\bf B519} 
(2001) 145, hep-th/0106191;
D. Klemm, hep-th/0106247.
\bibitem{BVV} J. de Boer, E. Verlinde and H. Verlinde,{\sl JHEP}
{\bf 08} (2000) 003, hep-th/9912012;
E. Verlinde and H. Verlinde, {\sl JHEP} {\bf 05} (2000) 034, 
hep-th/9912018.
\bibitem{SV} I. Savonije and E. Verlinde, hep-th/0102042.
\bibitem{nooqr} S. Nojiri and S.D. Odintsov, 
{\sl Class.Quant.Grav.} {\bf 18} (2001) 5227, hep-th/0103078;
Y.S. Myung, hep-th/0103241;
N.J. Kim, H.W. Lee and Y.S. Myung, hep-th/0104159.
\bibitem{others}
E. Verlinde, hep-th/0008140;
D. Kutasov and F. Larsen, hep-th/0009244;
F.-L. Lin, hep-th/0010127;
S. Nojiri and S.D. Odintsov, {\sl Int.J.Mod.Phys.} {\bf A16} 
(2001) 3273, hep-th/0011115;
B. Wang, E. Abdalla and R.-K.Su, hep-th/0101073
D. Klemm, A. Petkou and G. Siopsis, hep-th/0101076;
Y.S. Myung, hep-th/0102184;
R. Brustein, S. Foffa and G. Veneziano, hep-th/0101083;
R.-G. Cai, hep-th/0102113;
%
R.-G. Cai and Y.-Z. Zhang, hep-th/0105214; 
%
A. Biswas and S. Mukherji, hep-th/0102138;
D. Birmingham and S. Mokhtari, hep-th/0103108
D. Klemm, A. Petkou, G. Siopsis and D. Zanon, hep-th/0104141; 
D. Youm, hep-th/0105036, 0105093;
S. Nojiri, O. Obregon, S.D. Odintsov, H. Quevedo and M.P. Ryan, 
{\sl Mod.Phys.Lett.} {\bf A16} (2001) 1181, hep-th/0105052; 
R.-G. Cai, Y.S. Myung and N. Ohta, hep-th/0105070;
S. Nojiri, S.D. Odintsov and S. Ogushi, hep-th/0105117;
B. Wang, E. Abdalla and R.-K. Su, hep-th/0106086.
\bibitem{Cardy} J.L. Cardy, {\sl Nucl.Phys.} {\bf B270} 
(1986) 967. 
\bibitem{sta} A. Starobinsky, {\sl Phys.Lett.} {\bf B91} (1980) 99.

\bibitem{BBM} V. Balasubramanian, J. de Boer and D. Minic, 
hep-th/0110108. 
\bibitem{HHR2} S.W. Hawking, T. Hertog and H.S. Reall,
{\sl Phys.Rev.} {\bf D63} (2001) 083504, hep-th/0003052.
\bibitem{vilenkin} A. Vilenkin, {\sl Phys.Rev.} {\bf D32} (1985) 2511.
\end{thebibliography}
\end{document}